# Strong Anisotropy in Liquid Water upon Librational Excitation using Terahertz Laser Fields

Fabio Novelli[1], Luis Ruiz Pestana[2,3,4], Kochise C. Bennett[2,3], Federico Sebastiani[1], Ellen M. Adams[1], Nikolas Stavrias[5], Thorsten Ockelmann[1], Alejandro Colchero[1], Claudius Hoberg[1], Gerhard Schwaab[1], Teresa Head-Gordon[2,3,†], Martina Havenith[1,*]

[1]*Department of Physical Chemistry II, Ruhr University Bochum, 44780 Bochum, Germany*

[2]*Chemical Sciences Division, Lawrence Berkeley National Laboratory*

[3]*Pitzer Center for Theoretical Chemistry, Departments of Chemistry, Chemical and Biomolecular Engineering, and Bioengineering, University of California, Berkeley, Berkeley, CA 94720*

[4]*Department of Civil, Architectural, and Environmental Engineering, University of Miami, Coral Gables, FL, USA*

[5]*Radboud University, FELIX Laboratory, Toernooiveld 7, Nijmegen, Netherlands*

**Abstract**

Tracking the excitation of water molecules in the homogeneous liquid is challenging due to the ultrafast dissipation of rotational excitation energy through the hydrogen-bonded network. Here we demonstrate strong transient anisotropy of liquid water through librational excitation using single-color pump-probe experiments at 12.3 THz. We deduce a third order response of $\chi^3$ exceeding previously reported values in the optical range by three orders of magnitude. Using a theory that replaces the nonlinear response with a material response property amenable to molecular dynamics simulation, we show that the rotationally damped motion of water molecules in the librational band is resonantly driven at this frequency, which could explain the enhancement of the anisotropy in the liquid by the external Terahertz field. By addition of salt ($MgSO_4$), the hydration water is instead dominated by the local electric field of the ions, resulting in reduction of water molecules that can be dynamically perturbed by THz pulses.

[†]Correspondence to: thg@berkeley.edu

[*]Correspondence to: martina.havenith@rub.de

**INTRODUCTION**

Driving molecules in the gas phase has been possible for more than two decades using both static electric fields[1–3] and oscillating laser fields[4–6]. Terahertz-Kerr (THz-Kerr) techniques, that typically use resonant pump pulses at around 1 THz and an off-resonant optical probe[7–13], have shown initial success in enhancing the birefringence of liquid acetonitrile and biphenyl by resonant excitation of librational modes[14,15]. However, for bulk liquid water, exciting molecules using a laser source is still a challenge because the rotational motion of individual molecules is hindered by an extensive hydrogen-bonded network that can rapidly dissipate external excitations[16–19] as demonstrated by Cook *et al.*[20] who reported that the THz-induced excitation of water molecules in the liquid phase is sensitive to collective motions. Non-equilibrium molecular dynamics simulations have shown a rapid loss of the excitation energy of a single rotationally excited water molecule in the liquid phase on the sub-100 fs timescale, where the dominant dissipation pathway involves rotation of the hydrogen-bonded molecules in the first hydration shell of a central water[21]. This dissipation timescale is commensurate with previous experimental and theoretical studies showing that librational motions relax on a sub-200 fs timescale[21–26], with a characteristic decay time of 75 fs[27,28]. As a result, the third-order birefringence induced in water by either static fields[29,30] or laser excitation[13,31–39] has been found to be small, $3.1 \cdot 10^{-16} \frac{cm^2}{V^2}$ and $2.6 \cdot 10^{-18} \frac{cm^2}{V^2}$, respectively.

Here, we report single-color THz pump–THz probe experiments carried out on the ultra-bright free electron laser (FEL) at the Nijmegen FELIX laboratory that provides laser induced transient anisotropy in bulk liquid water, which is found to be orders of magnitude larger at 12.3 THz than previously reported values for excitations in the optical range. Using supporting theory that replaces the four point time correlation of the third order response with a material response property that can be simulated with molecular dynamics, we demonstrate that the birefringent enhancement is the result of resonance of the water modes at the lower frequency end of the librational band, colloquially referred to as "rocking" of the water dipole in the collective hydrogen-bonded network[40–43]. This resonant enhancement of the angular displacements of the water dipole compared to the ground state thereby increases the ability for increasing the anisotropy of the liquid through the applied electric field.

Furthermore, we would expect that the THz-induced anisotropy will be reduced if a locally stronger electric field is present, as in the case of hydration waters near the ions in a salt solution. Interestingly, our FEL experiments show that a 0.6 M MgSO₄ solution exhibits a significantly decreased birefringent response with respect to bulk water, and simulations of MgSO₄ solution at the same concentration also show a decreased signal of comparable magnitude. A detailed analysis of the simulation results reveals that the peak of the frequency spectrum of the water molecules in the hydration shell of the ions, in particular of $Mg^{2+}$, is blue-shifted with respect to bulk water. Therefore, the water molecules in the proximity of the ions don't contribute as much to the response when the system is excited at the resonant frequency of bulk water.

Our results show that the THz pump pulse creates a strong anisotropy in the bulk liquid, which is decreased upon addition of simple salt solutes. This paves the way for future THz pump-probe experiments, interpretable with a theory that yields tractable simulated observables using molecular dynamics, which can measure and/or alter the strength of anisotropic effects of water at biological and materials interfaces in a controlled way.

## METHODS

**Source.** The source emits "macropulses" with a repetition rate of 10 Hz. Each macropulse consists of 200 individual pulses with a temporal separation of 40 ns (25 MHz). The individual pulse length used was 3.6 ps full-width at half-maximum (FWHM) resulting in a pump-probe overlap of ~5 ps. When FELIX is tuned to emit radiation at 12.3 THz, and taking into account the reflection losses at the cell window, the typical fluence of each individual pump pulse was 3.1 mJ/cm$^2$, resulting in a peak power (field strength) of 900 MW/cm$^2$ (0.9 MV/cm) at the sample. The total macropulse power is estimated to 200x the peak power of each individual pump pulse, and amounts to 180 GW/cm$^2$. We used this large total power to estimate a lower limit for the birefringence from Eq. (1). The time zero calibration procedure, the spot sizes, and the effect of the pulse duration on the measured pump-probe signal are reported in the Supplementary Information (SI). All the experiments have been performed in the stable-focus mode of operation of FELIX, where no sub-pulses develop. As demonstrated previously[44,45], in this case all the micropulses have approximately the same frequency and temporal structure, and the FEL spectrum envelope corresponds to the short duration of the micropulse.

**Detection**. The detector is extrinsic photoconductive Germanium doped with Gallium, about 5 mm large, whereas the probe spot on the detector plane is ~1 mm. The detector is embedded in an Infrared Labs HDL5 cryostat with a 2 mm HDPE entrance window. The liquid sample thickness ($d = 25\ \mu m \sim 4$ times larger than then penetration depth[46–49]) is chosen to have an optimal probe voltage at the detector, in the middle of its dynamical range where it also displays a linear response. The output of the detector is integrated over all the pulses within a macropulse, and individual micropulses are not detected. For a typical signal see for example Fig.3c of Ref.[44]. After integration and amplification of the detector output, we used a 1 MHz low-pass filter at the input of the scope (2 GHz, 10 bit, 1 MΩ coupling) to suppress high frequency noise. The waveform we measure has two boxcars and takes two averages, one for the baseline and one for the signal. The pump-probe signal is obtained by averaging over the entire flat region of the macropulse, excluding the tails (see Fig.3c and Fig.6 in Ref.[44]). No further filtering and processing of the data is performed.

**Data analysis**. Figure 1 shows the experimental setup of the ultra-bright free electron laser FELIX. The beam is split into the pump and probe parts: the pump beam is vertically polarized (V) with respect to the optical table and the probe beam is polarized at an angle $\theta$ with respect to the pump direction (Fig. 1a). After the liquid sample, we rotate the polarizer (analyzer) to detect the vertical and horizontal components of the probe light. The resulting $T_V$ and $T_H$ transmissions are shown in Fig. 1b averaged over one macropulse, when the pump laser is blocked, and in Fig. 1c when the pump is on. Further details on the pump-probe setup at FELIX can be found in Ref.[44,45,50–52].

Our goal is to measure the peak magnitude of the third-order response $\chi^3$, which at a given pump-probe delay of $t = 0$ (see Eq. (4.1.20) in Ref.[53]) can be deduced from:

$$\chi^3 = \frac{\Delta n(0)}{I} \frac{n^2}{\left(283 \cdot \frac{V^2}{W}\right)} \tag{1}$$

where $I$ is the intensity, and $n$ is the index of refraction of the sample at equilibrium. In case of relaxation between subsequent micropulses, $I$ would correspond to the intensity of a single micropulse. However, since we observe an increase of our non-linear signal during the macropulse (see Fig.1c), we took $I$ as the power of the macropulse. $\Delta n$ was calculated as given in Eq. (2)[13,14,54]:

$$\Delta n(t) = n_V(t) - n_H(t) = \frac{\lambda \overline{\epsilon}(t)}{2\pi d^*} \tag{2}$$

where $n_V(t)$ and $n_H(t)$ are the index of refraction parallel and perpendicular to the pump polarization direction, respectively, $\lambda$ is the wavelength, and $d^*$ is the penetration depth of the probe radiation. The amplitude of the birefringence $\overline{\epsilon}(t)$ is approximately related to the birefringent signal $\epsilon(\theta, t)$ through[14]

$$\epsilon(\theta, t) = \overline{\epsilon}(t)\left(cos^3(\theta) - cos(\theta)\right) \tag{3}$$

where $\theta$ is the polarization of the probe beam.

At equilibrium, when the pump is blocked (Fig. 1b), the vertical (V) and horizontal (H) probe light intensities transmitted by the analyzer are $T_V(\theta) = e^{-\alpha d}cos^2\theta$ and $T_H(\theta) = e^{-\alpha d}sin^2\theta$, where $\alpha$ is the absorption coefficient at equilibrium, and $d$ the sample thickness. When the pump is active (Fig. 1c), the parallel and perpendicular absorption coefficients, $\alpha_V^*(t)$ and $\alpha_H^*(t)$, can change due to a change in the excited libration population. Furthermore, the probe can acquire a birefringent component $\epsilon(\theta, t)$, with $t$ the pump-probe delay. The pump-perturbed transmission coefficients averaged over the full macropulse, $T_V^*$ and $T_H^*$, and transmitted by the analyzer when the pump is on are given as:

$$T_V^*(\theta, t) = e^{-\alpha_V^*(t)d^* - \alpha(d-d^*)}cos^2\big(\theta + \epsilon(\theta, t)\big) \tag{4}$$

$$T_H^*(\theta, t) = e^{-\alpha_H^*(t)d^* - \alpha(d-d^*)}sin^2\big(\theta + \epsilon(\theta, t)\big) \tag{5}$$

where we assumed that the pump-induced transient signal originates from a sample with a thickness equal to one penetration depth, $d^* \approx 6 \ \mu m$ at 12.3 THz[46–49]. In these equations, we implicitly assumed that the pump-perturbed absorption coefficients $\alpha_V^*(t)$ and $\alpha_H^*(t)$ are, to a first approximation, independent of the probe polarization direction.

The pump-probe measurements were performed in balanced-detection mode[44,45,50–52]. We measure the difference between the intensity of the probe beam and the reference beam, which is a co-propagating duplicate of the probe delayed by 20 ns at each pump-probe delay $t$. The pump-probe signal is obtained by measuring simultaneously the probe and the reference pulse. In this way we can directly access the relative variation of the sample transmission and measure the pump-probe traces:

$$\frac{\Delta T_V(\theta, t)}{T_V(\theta)} = \frac{T_V^*(\theta, t) - T_V(\theta)}{T_V(\theta)} = e^{-\Delta\alpha_V(t)d^*}\frac{cos^2\big(\theta + \epsilon(\theta, t)\big)}{cos^2\theta} - 1 \tag{6}$$

$$\frac{\Delta T_H(\theta, t)}{T_H(\theta)} = \frac{T_H^*(\theta, t) - T_H(\theta)}{T_H(\theta)} = e^{-\Delta\alpha_H(t)d^*}\frac{sin^2\big(\theta + \epsilon(\theta, t)\big)}{sin^2\theta} - 1 \tag{7}$$

where $\Delta\alpha_V(t) = \alpha_V^*(t) - \alpha$ and $\Delta\alpha_H(t) = \alpha_H^*(t) - \alpha$ are the time-dependent absorption changes along and perpendicular to the pump polarization direction.

Thus, to estimate the third order response $\chi^3$ from an experiment at a full macropulse pump intensity $I$ and probe polarization $\theta$, we measure the pump-probe traces, $\frac{\Delta T_V(\theta, t)}{T_V(\theta)}$ and $\frac{\Delta T_H(\theta, t)}{T_H(\theta)}$, and then use Eq. (6) and (7) to solve for $\epsilon(\theta, t)$, $\Delta\alpha_V(t)$, and $\Delta\alpha_H(t)$. By measuring the pump-probe traces at two different probe-polarization angles and focusing just on the pump-probe delay around $t$=0 we can obtain an overdetermined system of linear equations that we can solve numerically to estimate $\epsilon(\theta)$, $\Delta\alpha_V$, and $\Delta\alpha_H$. Finally, by insertion of $\epsilon(\theta)$ into Eqns. (1), (2) and (3) we obtain $\chi^3$.

Since the pump-probe signal is increasing during the macropulse by a factor of ca 5x (Fig.1d), we estimate the intensity $I$ in Eq. (1) to the total power of the macropulse. In this way, we perform a conservative estimate of the third order response from liquid water.

**Comparison with Optical Kerr (OKE)**. It is important to note that our experiments are different compared to an optical Kerr[55–64] experiment as follows. When $\Delta\alpha_V(t)d^*, \Delta\alpha_H(t)d^*, \epsilon(\theta, t) \ll 1$ we obtain from the Taylor series expansion of Eqns. (6) and (7)

$$\frac{\Delta T_V(\theta,t)}{T_V(\theta)} \approx -\Delta\alpha_V(t)d^* - 2\epsilon(\theta,t)\tan\theta \tag{8}$$

$$\frac{\Delta T_H(\theta,t)}{T_H(\theta)} \approx -\Delta\alpha_H(t)d^* + 2\epsilon(\theta,t)\cot\theta \tag{9}$$

In the case that there is no absorption change ($\Delta\alpha_V = \Delta\alpha_H = 0$), as in a typical off resonant optical Kerr experiment, Eq. (8) and (9) coincide with equations reported previously[13,14,54], and the birefringence $\epsilon$ is related to the instantaneous response of the electronic system, i.e., to the induced dipole which stems from polarizability[14,59]. In the present case, the transient absorption terms $\Delta\alpha_V$ and $\Delta\alpha_H$ cannot be neglected, as both pump and probe are resonant with a sample excitation, and both absorption terms can be different than zero ($\Delta\alpha_V \neq 0; \Delta\alpha_H \neq 0$). As another extreme, ultrafast dichroism[22,23,27,28,65–68] represents a sub-class of resonant experiments where the pump-probe signal is dominated by anisotropic transient absorption ($\Delta\alpha_V \neq \Delta\alpha_H \neq 0$). In a typical dichroic experiment on an intramolecular vibrational mode of liquid water[23,27,28,69], the V-polarized pump pulse excites a fraction of molecules. In this case, the rotational anisotropy $R(t) = \left(\frac{\Delta T_V(t)}{T_V} - \frac{\Delta T_H(t)}{T_H}\right) / \left(\frac{\Delta T_V(t)}{T_V} + 2\frac{\Delta T_H(t)}{T_H}\right)$ is independent of the vibrational populations, and gives information on the time it takes for the fraction of pump-excited dipoles to diffuse away from the pump polarization direction[70]. If the rotational diffusion is extremely fast, $\frac{\Delta T_H}{T_H} \approx \frac{\Delta T_V}{T_V}$ and $R \rightarrow 0$; if it is infinitely slow[71,72], $\frac{\Delta T_H}{T_H} \approx \frac{1}{3}\frac{\Delta T_V}{T_V}$ and $R \rightarrow 0.4$. In this picture, the value of $\frac{\Delta T_H}{T_H}$ must be between $\frac{\Delta T_V}{T_V}$ and $\frac{1}{3}\frac{\Delta T_V}{T_V}$ at each pump-probe delay[23,27,28,69,70]. On the contrary, in a Kerr experiment the pump pulse additionally exerts a torque on the molecular dipoles[14]. The pump induces anisotropy, the sample becomes birefringent, and the signs of $\frac{\Delta T_H}{T_H}$ and $\frac{\Delta T_V}{T_V}$ are opposite[13,14,54,73]. For a summary of previous results see Table S1 in the SI.

**Sample**. Ultra-pure liquid water (Fig. 2, Fig. 3) and 0.6 M MgSO₄ aqueous solutions (Fig. 5) were loaded into a static cell with diamond windows. A Kapton spacer with thickness $d = 25 \, \mu m$ is placed in between the diamond windows. All the measurements were repeated multiple times to ensure reproducibility and avoid systematic errors. In order to keep the same sample thickness, the liquid cell was mounted and filled by flushing the liquid sample. Also, the sample position was kept the same for the different measurements: the cell fits into the thermally-stabilized copper plate with the aid of magnets. The first pulse in each macropulse interacts with liquid water which is stabilized at 20±0.05 °C by a recirculating chiller. However, during each macropulse, the train of 200 pulses can induce a temperature increase in the sample. An upper limit for the sample temperature during each macropulse can be estimated as ~40 °C (diffusive heating). Between subsequent macropulses, the temperature decreases to 20 °C within ~4.5 ms due to diffusion limited cooling in the temperature stabilized sample cell[74,75]. Details are given in the SI.

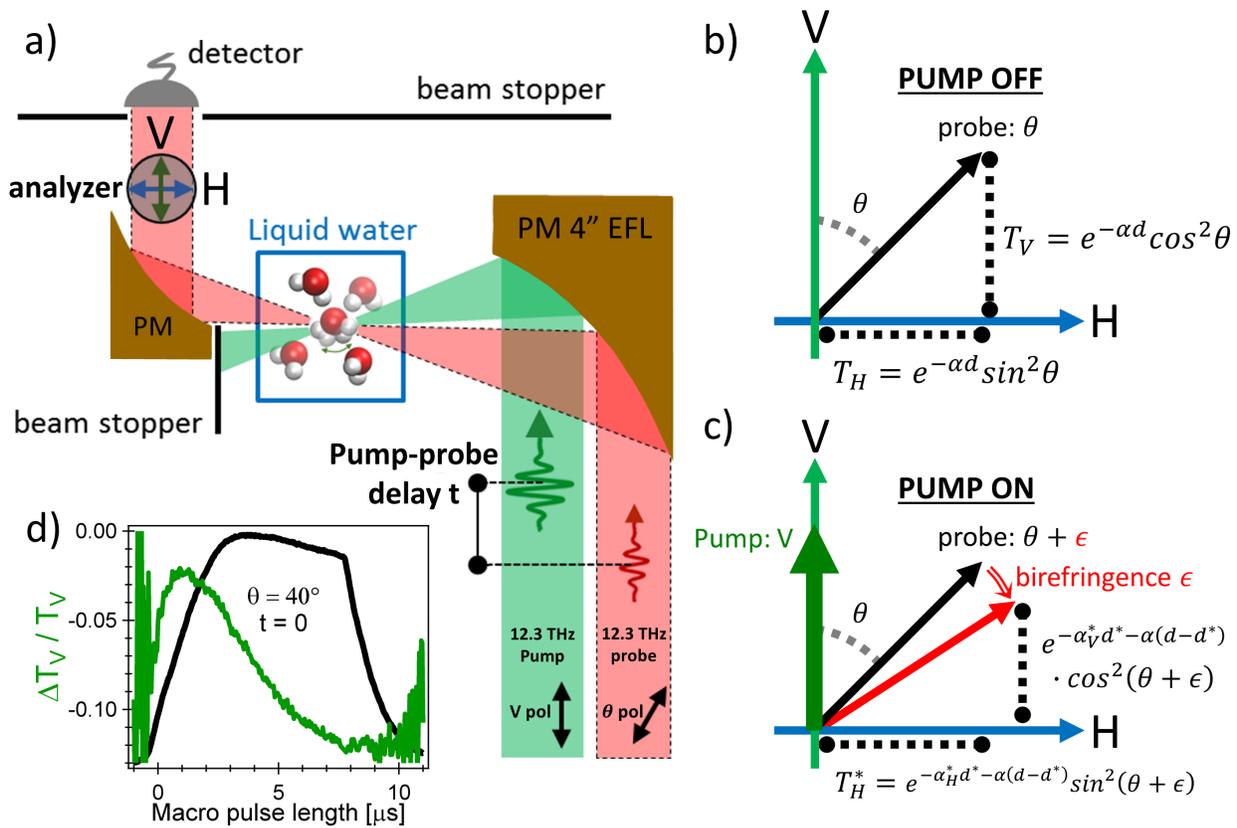

**Figure 1.** Experimental configuration of the THz pump-probe experiments using the FELIX free electron laser. The beam is split into a pump (green) and a 70x weaker probe (red) component. A 10.16 cm (4") diameter off-axis parabolic mirror (PM) with 10.16 cm effective focal length (EFL) focuses both pump and probe beams at the sample position. The temporal overlap was found by placing a 200 μm metal pinhole at the sample position (see Fig. S1 in the Supporting Information and the description therein). The liquid sample is loaded into a static liquid cell with diamond windows. After the sample, a second PM with 12.7 cm EFL collects radiation from the probe beam. A polarizer (analyzer) with $10^4$:1 extinction ratio is placed before the detector to allow detection of either the vertical, $T_V$ (green), or horizontal, $T_H$ (blue), components of the probe transmission. d) In black we show the transmission of the probe beam at the temperature $\overline{T}$ = 20 °C from the onset of the macropulse (0 μs) to the end of the macropulse (ca. 8 μs). In green we show the corresponding relative variation of the transmission $\frac{\Delta T_V(\theta,t)}{T_V(\theta)}$ at pump-probe overlap t = 0 ps and probe polarization θ=40°, see Eq. (6). The magnitude of the pump probe signal increases during the macropulse.

**Molecular dynamics.** We performed molecular dynamics simulations using the TINKER molecular dynamics package[76] and the polarizable force field AMOEBA, specifically the AMOEBA14 water model[77]. We study two different systems: a pure water system with 512 water molecules packed at a density of 1 g/cm³, and a MgSO₄ solution with 975 water molecules and an salt concentration of 0.5 M (10 Mg²⁺ and 10 SO₄²⁻ ions). All the simulations are carried out in the NVT ensemble at room temperature using a Nose-Hoover thermostat with a coupling constant of 0.1 ps and a Beeman integrator. The timestep in all the simulations is 0.5 fs. Each trajectory has a length of 25 ps, enough to sample the ultra-fast rotational modes of interest. The oscillating external field is implemented by adding a component to the permanent electrostatic field that changes in time according to a sinusoidal function, i.e. E=A sin($\omega_{pump}$ t), where A is the amplitude of the electric field and $\omega_{pump}$ is the frequency. We performed simulations at $\omega_{pump}$ = 1.0, 7.8, 12.3, and 200 THz, at three different amplitudes of the electric field 1 MV/cm, 2 MV/cm, and 3 MV/cm, and for 3 different polarization directions. The results shown in the all the figures correspond to the average among the different polarization directions. From the simulation trajectories we calculate the angle $\varphi_i(t)$ between the electric field $\vec{E}$ and the dipole moment of the $i$-th water molecule $\vec{\mu}_i$ at time $t$. From these time series of the orientation angles it is trivial to calculate $\langle \frac{d}{dt}[cos\varphi(t)] \rangle$, and hence $S(\omega)$ using Eq. (20). The angular brackets indicate the average over all the molecules in the system in the canonical ensemble. In order to validate the AMOEBA force field, we compared the $S(\omega)$ of pure water from two different force-field water models (AMOEBA14 and TIP4P-Ew) to the results from ab initio molecular dynamics (AIMD) based on density functional theory and the functional B97M-rV (see the SI). We find that while AMOEBA (the polarizable force field) is in good agreement with the AIMD results, the classical force field with fixed charges TIP4P-Ew strongly disagrees.

**RESULTS**

**Measurements at 12.3 THz**. We carried out the single-color THz pump–THz probe experiments at 7.8, 12.3, and 13.9 THz using the experimental setup and analysis described above. The single-color pump-probe experiment at 12.3 THz for probe polarization angles of $\theta$=40° and $\theta$=50° measures a negative vertical (Fig. 2a) and positive horizontal (Fig. 2b) pump-probe trace around pump-probe delay $t$=0, which indicates that the probe light polarization acquires ellipticity at pulse overlap, and thus that the interaction between water and the pump pulses induces an ultra-fast ($\ll 5\ ps$) anisotropy in the bulk liquid. From the difference between the horizontal pump-probe traces measured at $\theta$=40° and $\theta$=50° we estimate an anisotropic signal of no less than 1% (compare light and dark blue curves in Fig. 2b). Furthermore, the maximum modulus of the signal for vertical (Fig. 2a) and horizontal (Fig. 2b) pump-probe traces differs, which implies that the pump-probe signal is partially dichroic ($\Delta\alpha_V(t) \neq \Delta\alpha_H(t) \neq 0$). The probe ellipticity contains effects from both dichroism and the birefringence induced in the bulk liquid by the pump pulses.

The peak values for $\frac{\Delta T}{T}$ measured on liquid water solutions are approximately $-10\%$ for vertical pump-probe traces (see Fig.2a) and $+2\%$ for the horizontal (Fig.2b). In case of diffusive heating[46,48] we expect an isotropic and negative $\frac{\Delta T}{T}$ signal. The measured pump-probe signal is anisotropic and cannot be attributed solely to heating. Thus, we can exclude diffusive heating as an explanation for the experimentally determined $\frac{\Delta T_H}{T_H} > 0$.

We measure the vertical and horizontal pump-probe traces and use Eqns. (6)-(7) to solve for $\epsilon(\theta,t)$, $\Delta\alpha_V(t)$, and $\Delta\alpha_H(t)$. From $\epsilon(\theta)$ and Eqns.(1)-(3) we calculate the third-order birefringence $\chi^3 \approx 1.3 \cdot 10^{-15} \frac{cm^2}{V^2}$ at pump-probe overlap. From the peak pump-probe values obtained in pure water (Fig. 2), we also obtain the absorption changes:

$$\Delta\alpha_H = 16 \pm 3.6\ cm^{-1} \text{ (averaged over } \theta = 40° \text{ and } \theta = 50°) \tag{10}$$

$$\Delta\alpha_V = 132 \pm 4.2\ cm^{-1} \text{ (averaged over } \theta = 40° \text{ and } \theta = 50°) \tag{11}$$

These transient absorption terms at pulse overlap are different and larger than zero, and quantify both the amount of dichroic pump-probe signal as well as the change in absorption due to the pump-induced anisotropy in the bulk liquid.

We checked the assumption that the transient absorption coefficients $\Delta\alpha_V$ and $\Delta\alpha_H$ are independent on the polarization direction of the **probe**. We performed additional pump-probe measurements at pulse overlap $t = 0\ ps$ and for probe polarizations $\theta$=0° and $\theta$=90°. We measured the vertical pump-probe trace $\frac{\Delta T_V(\theta=0°,t=0\ ps)}{T_V(\theta=0°)} = (-87 \pm 3) \cdot 10^{-3}$ as well as the horizontal signal $\frac{\Delta T_H(\theta=90°,t=0\ ps)}{T_H(\theta=90°)} = (+3 \pm 3) \cdot 10^{-3}$. For $\theta$=0°, Eq. (8) becomes $\frac{\Delta T_V(\theta=0°,t=0\ ps)}{T_V(\theta=0°)} = (-87 \pm 3) \cdot 10^{-3} \approx -\Delta\alpha_V d^*$. For $\theta$=90°, Eq. (9) is $\frac{\Delta T_H(\theta=90°,t=0\ ps)}{T_H(\theta=90°)} = (+3 \pm 3) \cdot 10^{-3} \approx -\Delta\alpha_H d^*$. From these equations we deduce

$$\Delta\alpha_H = -5 \pm 5\ cm^{-1} \text{ for } \theta = 90° \tag{12}$$

$$\Delta\alpha_V = 145 \pm 5\ cm^{-1} \text{ for } \theta = 0° \tag{13}$$

which agree within three standard deviations with the values reported in Eqns.(10)-(11) assuming $\Delta\alpha_H \neq \Delta\alpha_H(\theta)$ and $\Delta\alpha_V \neq \Delta\alpha_V(\theta)$.

We can now estimate the error bar for the third-order birefringence observed in liquid water at 12.3 THz. We forcefully insert and lock the values of $\Delta\alpha_H \approx -5\ cm^{-1}$ and $\Delta\alpha_V \approx 145\ cm^{-1}$ (Eqns. (12)-(13)) into Eqns. (6)-(7). Together with the experimentally measured pump-probe signals reported in Fig.2, we obtain a lower limit for the third-order response associated to birefringence, $\chi^3 \approx 0.8 \cdot$ $10^{-15} \frac{cm^2}{V^2}$. Thus, we deduce a birefringence of $\chi^3 = (1.3 \pm 0.5) \cdot 10^{-15} \frac{cm^2}{V^2}$. This value is about one order of magnitude larger than for static fields[29,30] ($3.1 \cdot 10^{-16} \frac{cm^2}{V^2}$) and exceeds by three orders of magnitude the nonlinear response of pure water at optical and near-infrared[34,35,37–39,78–81] frequencies, $2.6 \cdot 10^{-18} \frac{cm^2}{V^2}$. We note that the third-order susceptibility of water in the gas phase has been measured previously[53,82–84] and amounts to $\chi^3 \lesssim 3 \cdot 10^{-21} \frac{cm^2}{V^2}$. This is 6 orders of magnitude smaller than the third-order response we detect, $\chi^3 = (1.3 \pm 0.5) \cdot 10^{-15} \frac{cm^2}{V^2}$. Thus, we cannot attribute the enhancement of $\chi^3$ compared to those at higher frequencies to the formation of water vapor. Furthermore, we want to point out that we assumed an additive effect of the subsequent micropulses ($I$ was taken as $I$ of the entire macropulse). Therefore, the deduced value corresponds to a lower limit.

**Control experiments**. The pump-probe signal scales with the pump intensity and is probe-independent (see the SI), as expected for a $\chi^3$ process involving two pump fields. This observation suggests that high order nonlinearity, which could be polarization-dependent and display strong anisotropy[85], might not be the dominant term in our case. As detailed in the experimental geometry setup reported in Fig.1a, pump and probe beams are non-collinear, and the detector collects only radiation along the propagation direction of the probe beam. We checked that no pump scatter is detected. When the probe is blocked and the pump beam is opened, no scatter reaches the detector. Additional experiments performed with only one micropulse yielded nonconclusive results.

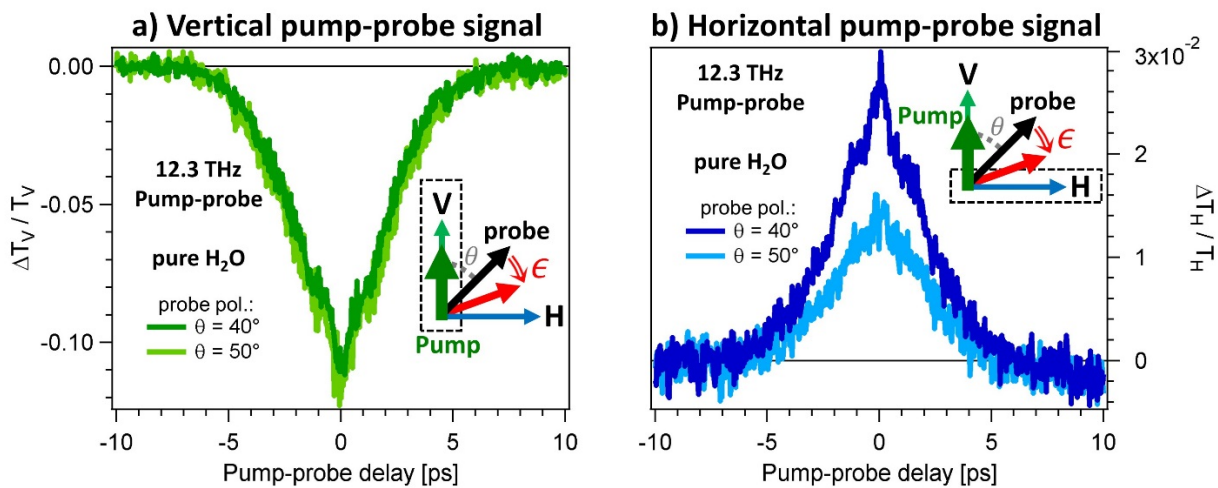

**Figure 2.** Polarization-dependent pump-probe of bulk liquid water at 12.3 THz. The pump-probe traces are recorded when the analyzer is set to the vertical (a) or horizontal (b) position. The probe polarization angle is set to 40° (dark green in panel (a); dark blue in (b)) or 50° (light green in panel (a); light blue in (b)) with respect to the pump laser (V). The noise level is less than ±2·10⁻³. The displayed transient signal is the average over the macropulse.

**Single color pump-probe at 7.8 and 13.9 THz on water**. In addition, we carried out a similar single-color pump-probe experiments at 7.8 and 13.9 THz with the probe polarization set to $\theta$=45° with respect to the vertically-polarized pump (Figure 3). At 13.9 THz, the pump energy deposited in the librational band of liquid water should be similar to that at 12.3 THz[46–49,86,87], and one would expect similar transient changes as described by $\Delta\alpha$ and $\Delta n$. We in fact observe a qualitatively similar signal at 13.9 THz compared to the 12.3 THz result, with a negative vertical pump-probe trace and a positive horizontal pump-probe signal at pulse overlap (comparing blue and green curves in Figures 3a and 3b). However, the penetration depth of liquid water[46–49] at 13.9 THz is reduced ($d^* \approx 4\ \mu m$) when compared to the penetration depth at 12.3 THz ($d^* \approx 6\ \mu m$), and consequently the pump-probe signals are expected to be smaller. However, the measurement noise is ~5x larger at 13.9 THz.

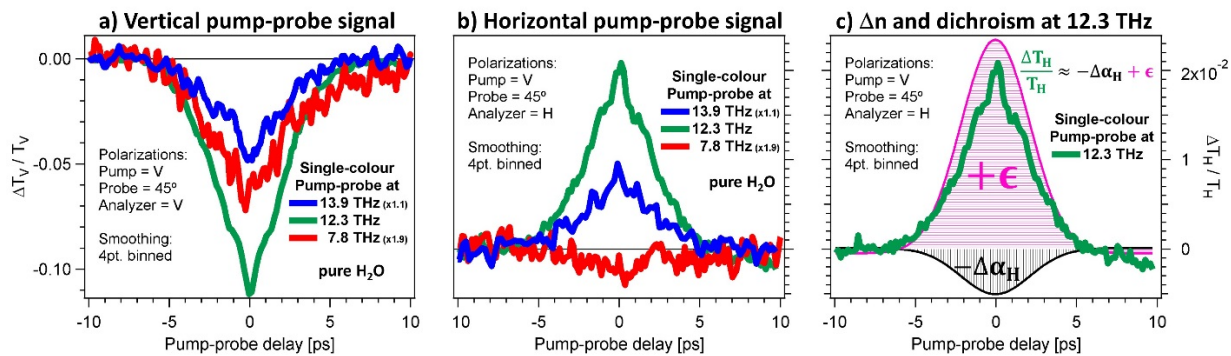

**Figure 3**. Frequency-dependent pump-probe signal in water. In panels (a) and (b) we show the transient response acquired with the polarizer set to pass vertically- or horizontally-polarized light, respectively. The single-color measurements are performed in pure water at 13.9 THz (blue), 12.3 THz (green), and 7.8 THz (red). The probe polarization is set to $\theta$=45° in all these measurements. The pump-probe trace at 13.9 THz (7.8 THz) is multiplied by a factor x900/800 (x900/470) to account for the different source brilliance. The measurements are noisier at 13.9 THz and 7.8 THz due to atmospheric absorption[88,89], thus all the data shown here are re-binned: every 4 consecutive points are replaced with their average. This reduces the temporal resolution from ~35 fs to ~140 fs. (c). From Eq. (9) we can visually separate the horizontal pump-probe trace at 12.3 THz shown in panel (b) into contributions from the birefringence (purple) and the transient absorption (black). The purple and black curves are guides to the eyes. Panels (b) and (c) share the same y-axis.

At 7.8 THz the absorption coefficient is reduced and the experimental noise is increased compared to the measurements at 12.3 THz. But in addition, at 7.8 THz the laser energy is deposited into both the intramolecular hydrogen-bond stretch as well as into a librational mode[47]. The excitation of the intermolecular hydrogen-bond stretch is not expected to lead to any notable anisotropy[14,15,47,90]. Thus, we expect a decrease of the anisotropy for pump-probe signals at 7.8 THz compared to 12.3 THz, as confirmed by the experiment (red vs. green curves in Figures 3a and 3b). The small negative horizontal pump-probe signal observed at 7.8 THz is consistent with a fast relaxation process, as expected for the hydrogen-bond stretch[47,90]. We propose that an increase in $\Delta\alpha_H$ (negative black signal in Fig. 3c) overcompensates a smaller birefringent signal expected at 7.8 THz, resulting in a net negative signal at 7.8 THz (red in Fig. 3b).

## DISCUSSION

1. **Possible origins of the signal increase over the macropulse**.

We demonstrated that pulses at 12.3 THz induce anisotropy in bulk water (Fig.2, Fig.3). The anisotropy is short-lived and lasts only during pump-probe overlap ($\ll 5\ ps$), but it also accumulates during the macropulse ($8\ \mu s$, Fig.1d). This unusual result could stem from the combination of several complex phenomena including nonlinear, thermal, and dissipative effects.

**Thermal grating**. The interference of the pump and probe pulses at spatial and temporal overlap can generate a thermal grating[91]. One effect of this grating is to diffract part of the intense pump beam into the direction of the probe[91,92]. In this way, more pump light reaches the detector at pump-probe overlap, and the relative variation of the transmission is positive. However, this is not in agreement with our observations, which yield a decrease, i.e. $\frac{\Delta T_V(t=0)}{T_V} \approx -10\%$ when the analyzer is aligned with the pump polarization direction (Fig. 2a). We also note that the temperature dependence of the refractive index of liquid water is smaller at 12.3 THz, and larger both at 7.8 THz as well as 13.9 THz[46]. If the transient signal would be dominated by a thermal grating, it should be larger at 7.8 and 13.9 THz, and smaller at 12.3 THz. However, the opposite is observed experimentally (Fig. 3).

**Thermal lensing**. When a medium is excited by a train of laser pulses, heat can accumulate along the spatial profile of the laser beam[93]. This generates a thermal gradient and, if the index of the refraction ($n$) of the material depends on the temperature ($\overline{T}$), the medium can act as a lens. The dioptric power of such a thermal lens is approximately[93,94] $f^{-1} = \frac{dn/d\overline{T}}{2kA}P$, where $f$ is the focal length, $k$ thermal conductivity, $A$ irradiated surface, and $P$ absorbed power transformed into heat under thermodynamic equilibrium. To first approximation, the focal length of the thermal lens should also be proportional to the fluence ($\Phi \propto f^{-1} \propto P$). Any thermal lensing would cause an additional focusing of the pump during the macropulse ($8\ \mu s$), which would increase the measured third-order nonlinear signal. We note that additional complex lensing phenomena, including non-linear and non-thermal effects such as filamentation and micro-lensing[95,96], are also possible.

**Acoustic effects.** In previous studies a propagation of optical phonon-like THz modes were found to be supported by the H-bond network[40]. The energy transfer depends on the group velocity of these propagating transverse waves. Predicted group velocities for wave propagation in the low frequency range vary between[40,97,98] 400 and 2500 m/s. The time it takes for a thermal lens to form (rise time, $r$) can be estimated to $r = \frac{w}{s}$, where $w$ is the laser beam waist and $s$ is the speed of sound in the liquid[99]. Considering the THz pump spot size (700 $\mu m$) and the group velocity, we estimate a rise time for the formation of a thermal lens of $r = 0.3 - 1.8 \, \mu s$. Thus, we would expect an underlying slower change in the signal due to the formation of thermal lensing which could cause improved focusing of the pump, and thus increase the measured third-order nonlinear signal during the macropulse (8 $\mu s$).

**Out of equilibrium dynamics**. It is not always possible to assign an effective time-dependent temperature to a system that is perturbed by an ultra-short light pulse. For some time during and following the light-matter interactions, in fact, the perturbed population of the (directly or indirectly) excited transitions is inconsistent with a Boltzmann distribution and the temperature is ill-defined[100,101]. The rise and decay times of a thermal component of a pump-probe signal depends on the details of the transitions which are initially photo-excited, how these are coupled to bath modes, whether or not the occupation of these low-frequency excitations resembles a thermal distribution[100,101] and, finally, how the probed transitions are affected by all this. Previous nonlinear experiments on liquid water reported about the "hot ground state" (HGS). The HGS is typically reached within 1 ps, and the response of the liquid resembles the one of heated water. However, the resemblance is misleading, because the HGS is not in thermal equilibrium. De Marco *et al.*[100] estimated that water is in "*a highly non-equilibrium state in which the population of eigenstates is not a Boltzmann distribution*" after the laser excitation. For these reasons, in the full macropulse experiments, the next pulse could interact with a non-relaxed liquid that is far from equilibrium. Thus, the increase of the magnitude of the pump-probe signal during the macropulse time (Fig.1d) could be related to the non-relaxed population of the librational mode. We speculate that in this case the anisotropic signal could be increased since the populated librational states are above the barrier for tunneling motion. In fact, bifurcational tunneling in the excited librational mode has been found to be increased by three orders of magnitude[102].

## 2. Theory

Next we turn to other independent assessments of the interpretation regarding the strong anisotropy in the librational region: (1) theory that provides a means to simulate the frequency spectrum of the rotational motion of bulk water molecules, which we find to be in excellent agreement with the THz pump-probe experiment. and (2) an additional 12.3 THz pump-probe experiment for a low concentration

MgSO₄ solution in which the absorption (and thus the heating effect) are the same within 1% of the bulk water value[103]. We observe a significant decrease of the signal by 16 (±7) % in the transient anisotropy of the MgSO₄ solution, supported by additional simulations, in line with the expectation of an increased third-order response.

**Theoretical Model**. To gain insight into the molecular mechanisms responsible for the extremely large experimental birefringent response, direct simulation of the pump-probe experiment requires the acquisition of a four-point correlation function in the time-domain, a task that remains prohibitively expensive for all but the simplest systems[24]. For this reason, we derive an approximation to the $\chi^3$ experimental response by a simpler time evolving variable that can be calculated using molecular dynamics simulations. To show this, we begin with the frequency-dependent expression for a heterodyne-detected signal:

$$S(\omega) = \Im\left[E_S^*(\omega)\langle\widehat{P}(\omega)\rangle\right] \tag{14}$$

where $\Im$ stands for the imaginary part, $E_S^*(\omega)$ is the complex conjugate of the probing electric field envelope, and $\widehat{P}(\omega)$ is the material polarization operator in the frequency-domain that contains field-matter interactions to arbitrary-order. Rigorously, pump-probe experiments require that we expand the material polarization operator $\widehat{P}$ to 3rd order in the field-matter interaction. However, if the perturbative expansion of the polarization is performed, a quantum model of the material excitation is required whose direct calculation is prohibitively costly for all but the simplest systems[24]. In order to bypass this issue, we explicitly add the oscillating electric field of the pump to the Hamiltonian for the time propagation. We then expand $\widehat{P}(\omega)$ to 1st order in the probe field, and obtain

$$\langle\widehat{P}(\omega)\rangle = \langle\hat{\mu}(\omega)\rangle = -\frac{i}{\hbar}\int dt\, e^{i\omega t}\int dt'\, E(t-t')\, Tr([\hat{\mu}(t'),\hat{\mu}(0)]\rho(0)) \tag{15}$$

where $\hat{\mu}$ is the dipole operator, so that the first equality is just the dipole approximation and $\rho$ is the material density matrix. We have replaced the full vector electric field by its scalar amplitude $E$ so that $\hat{\mu}$ is the projection of the dipole along the electric field. We assumed that the intensity envelope of the probe field varies slowly with respect to the coherence time between the two probe-matter interactions, and that the probe-matter interaction operators are in the Born-Oppenheimer approximation. Substituting the spectral envelope for the temporal E-field allows integration over $t$, giving

$$\langle\widehat{P}(\omega)\rangle = -\frac{i}{\hbar}E(\omega)\int dt'\, e^{i\omega t'}\, Tr([\hat{\mu}(t'),\hat{\mu}(0)]\rho) \tag{16}$$

The key quantity here is therefore the Fourier transform of the material response function, $J(t) = FT^{-1}[S(\omega)]$:

$$J(t) = -\frac{i}{\hbar}Tr([\hat{\mu}(t'),\hat{\mu}(0)]\rho) = \frac{i}{\hbar}Tr([\hat{\mu}(t'),\rho]\hat{\mu}(0)) \tag{17}$$

where we have used the cyclic invariance of the trace. The Fourier transform of this quantity gives the frequency domain response of the material system and thus contains all information about the light-matter interaction within this approximation. In the classical limit (the proof is given in the SI) we can write Eq. (17) as:

$$J(t) = \mu^2 \langle cos\theta(0) \frac{d}{dt}(cos\theta(t)) \rangle \tag{18}$$

where $\theta$ is the angle between the electric field and the molecular dipole. The integration over all initial conditions with respect to arbitrary time origins leads to:

$$J(t) = \langle \frac{d}{dt}(cos\theta(t)) \rangle \tag{19}$$

which using the identity $J(t) = FT^{-1}[S(\omega)]$ we can finally rewrite as:

$$S(\omega) = FT \left[ \langle \frac{d}{dt}(cos\theta(t)) \rangle \right] \tag{20}$$

This quantity, $S(\omega)$, can be computed using molecular dynamics simulations and includes the effect of multiple pulse interactions (macropulse).

**Simulation Results.** We performed simulations with the polarizable force field AMOEBA14 water model[77], which includes a many body electronic response of mutual polarization in addition to permanent multipoles. The model has been previously validated against the water THz spectrum[104] and *ab initio* molecular dynamics simulations using different functionals[104–106]. All computational details are described in the Methods section and in the SI.

At equilibrium (i.e. no pump or zero-field), the frequency spectrum of the rotational motion of bulk water molecules, $S(\omega)$ in Eq. (20), exhibits a relatively broad peak at 12-20 THz (Fig. 4). It is worth noting that the theoretical frequency spectrum shown in Fig. 4b is on the lower end of the librational band in standard IR experiments[46–49,86], which peaks at ~21 THz and contains other dynamical modes[47]. In this case the theory predicts that the underlying natural dynamics of damped dipole reorientations within the hydrogen-bond network are excited in the THz pump-probe experiment, inducing water dipole reorientations that span a large angular displacement on ultra-fast timescales of ~30-120 fs.

This is confirmed when we apply an oscillating electric field in the simulations and characterize the response of the system, for which we find that the spectrum $S(\omega)$ yields a much larger resonant response at $\omega_{pump}$ = 12.3 THz pump frequency than at $\omega_{pump}$ =1.0, 7.8, or 200 THz (Fig. 4d). As illustrated in Figure 4c, we define the relative peak intensity, $I_{rel}$, as the difference between the amplitude of $S(\omega)$ and the background spectral density at $\omega_{pump}$. Figure 4e shows the expected quadratic dependence of $I_{rel}$ on the field strength. Thus the simulated results are in excellent concordance with the THz pump-probe experiments, and provide a molecular interpretation that the large third order response in the bulk liquid arises from resonance of the THz pulse with the softer librational modes that create a more anisotropic liquid.

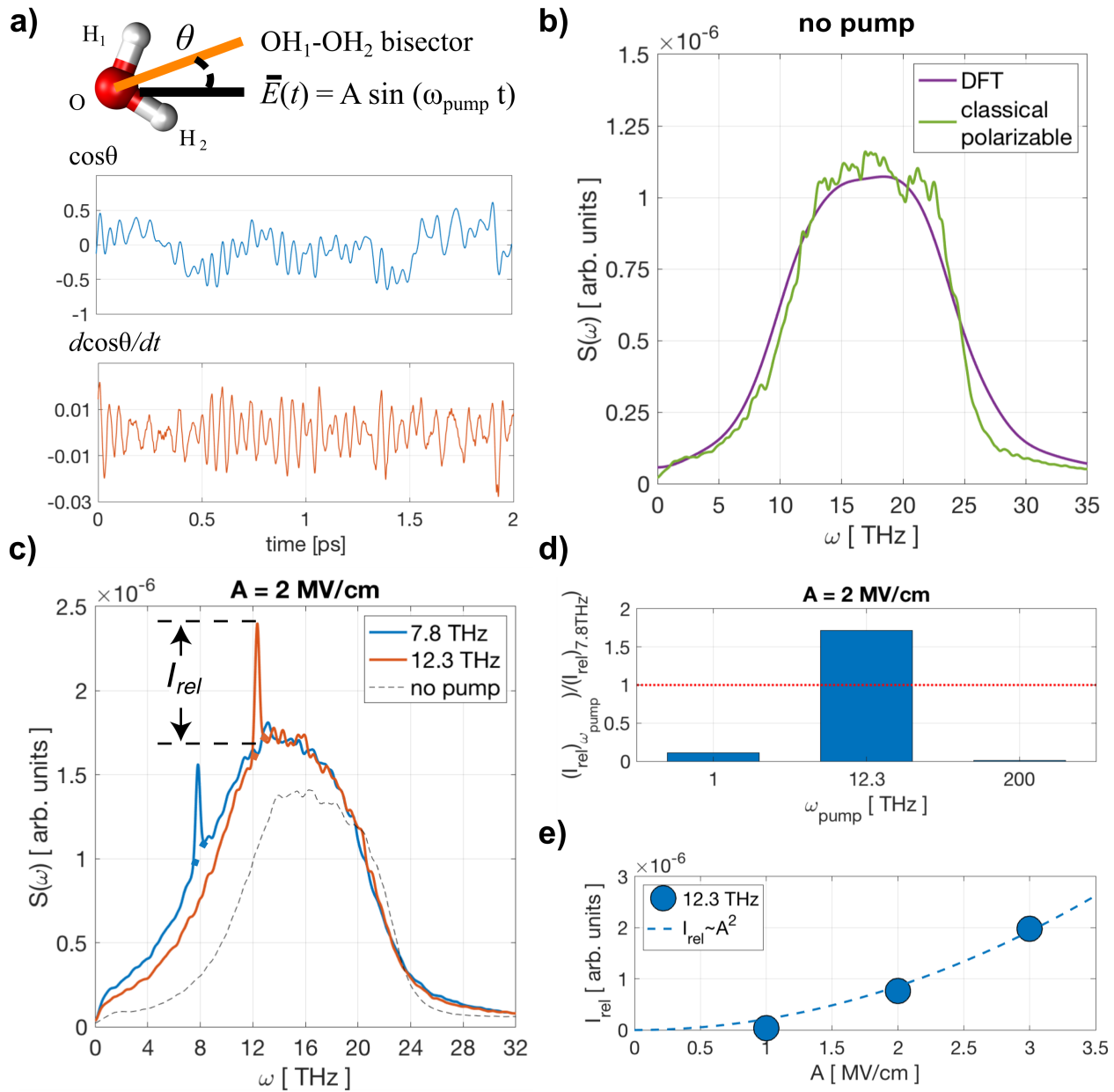

**Fig. 4.** Results from molecular dynamics simulations of water. (a) An example of the time evolution of $\theta(t)$ and $d\cos\theta/dt$ in a 2 ps time-interval and (b) $S(\omega)$ for a system with 64 water molecules calculated from Eq. (20) in the absence of an applied electric field; the polarizable force field AMOEBA[77] agrees well with the results from ab initio molecular dynamics using the B97M-rV density functional[105]. (c) $S(\omega)$ for systems with 512 water molecules in the presence of an applied oscillating electric field of 2 MV/cm at 7.8 and 12.3 THz. The definition of the relative peak amplitude, $I_{rel}$, is illustrated. (d) Ratio between $I_{rel}$ for electric fields applied at frequencies at 1.0, 12.3, and 200 THz with respect to the $I_{rel}$ at 7.8 THz. Similar to the experiments, a larger effect is observed at 12.3 THz, and not for other frequencies. (e) Dependence of $I_{rel}$ as a function of the amplitude of the applied electric field with a frequency of 12.3 THz.

**3. Theoretical and experimental benchmark on salt solution**. Our second piece of evidence that the pump-probe signal in the 12-20 THz range arises from strong anisotropy comes from additional pump-probe experiments at 12.3 THz for a low concentration $MgSO_4$ solution in which the absorption (and thus the heating effect) are the same within 1% of the bulk water value[103]. If the signal were to originate exclusively from heating, the response by pure water and a salt solution should be identical since the equilibrium absorption coefficient of the salt solution, specifically the 0.6 M $MgSO_4$ aqueous solution that we study, is only ~1% different than liquid $H_2O$ at 12.3 THz[107]. However, the results of the pump-probe measurements of a 0.6 M $MgSO_4$ aqueous solution at 12.3 THz (Fig. 5a-e) show a statistically significant drop of the signal at pump-probe overlap, $-16 \pm 7\%$ (as estimated by Gaussian fits), when compared to bulk water in line with the expectations from additional simulations.

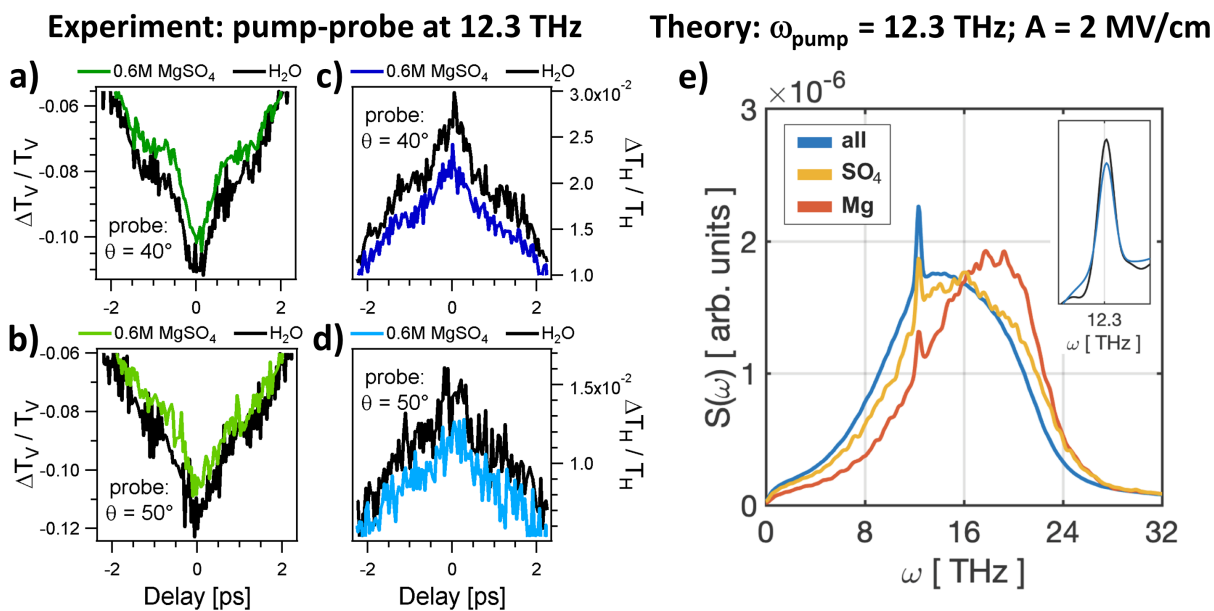

**Figure 5.** Polarization-dependent pump-probe of 0.6M $MgSO_4$ aqueous solution at 12.3 THz. We report in panels (a)-(b) and (c)-(d) the vertical, $\frac{\Delta T_V(\theta,t)}{T_V(\theta)}$, and horizontal, $\frac{\Delta T_H(\theta,t)}{T_H(\theta)}$, pump-probe signals, respectively. The probe polarization is set to $\theta$=40° (dark green in panel (a); dark blue in panel (c)) and to $\theta$=50° (light green in panel (b); light blue in panel (d)) with respect to the pump polarization direction (V). The pump-probe signals measured in pure water are shown in black lines for comparison. All the pump-probe signals are ~16% smaller in the salt solution. The data are not re-binned. All panels share the same time delay axis. The complete pump-probe traces are reported in Fig. S4 in the SI. (e) Results from molecular dynamics simulations of a 0.5 M $MgSO_4$ solution in the presence of an applied oscillating electric field of 2 MV/cm at 12.3 THz. The $S(\omega)$ arising exclusively from the water molecules in the first solvation layer of Mg and $SO_4$ ions is shown in red and yellow, respectively. The full $S(\omega)$ of the salt solution is shown in blue. In the inset we compare the full $S(\omega)$ of the $MgSO_4$ solution (blue) to the one of bulk water (black). The decreased response of water molecules in the salt solution with respect to bulk water is clear.

To determine the origin of the reduction in the pump-probe signal in the presence of salt we have performed molecular dynamics simulations of a comparable (0.5 M) $MgSO_4$ solution and calculated the frequency spectrum $S(\omega)$ of all the water molecules, as well as the $S(\omega)$ of just the water molecules in the hydration shell of the $Mg^{2+}$ and $SO_4^{2-}$ ions (Fig. 5e). The strong interaction between water and the ions, in particular with the $Mg^{2+}$ cation, manifests as a strong blue shift of ~8 THz of the resonant frequency, which effectively moves the response of the water in the hydration of $Mg^{2+}$ outside the resonant experimental frequencies. In the inset of Fig. 5e we compare the $S(\omega)$ of MgSO$_4$ solution (blue) and pure water (black) focusing on the response around the resonant peak at 12.3 THz. The comparison reveals a reduction in the birefringent third-order response (as quantified by $I_{rel}$) of $-20 \pm 3.5\%$, which is in good agreement with the experiment ($-16 \pm 7\%$). The strong electrostatic interaction between the divalent ions and water shifts the natural frequency of the rotational motion of the water molecules in the solvation shell cage towards higher frequencies. Therefore, when pumping the salt solution at the resonant frequency of bulk water molecules (e.g. 12.3 THz), the birefringent signal is decreased because the stiffened librational mode of the hydration water is out of the range of that frequency.

**CONCLUSION**

We have presented single-color pump-probe experiments at 7.8, 12.3, and 13.9 THz to measure the induced transient anisotropy in the bulk liquid phase. We estimate a third-order $\chi^3$ anisotropy of pure water at 12.3 THz that could exceed by three orders of magnitude that found for optical and near-infrared fields. We have presented new theory in which the four-point time correlation function, that typically would be required to simulate the nonlinear birefringent response, can be replaced with a formalism that implicitly represents the pump while only needing the two-point time correlation at the resulting probe frequency. Based upon a combined experimental and theoretical study we suggest that the pump-probe experiment at the 12.3 THz laser frequency could be resonant to the reorientation timescale of water dipoles in their hydration cages, and that application of an external field at this frequency might lead to transient anisotropy in a homogeneous liquid phase. The experimentally observed pump induced anisotropy at 13.9 THz is qualitatively similar. At 7.8 THz we observe a decreased transient signal. The decrease in signal magnitude is also observed in the simulation and is attributed largely to the fact that the timescale of the driving field is detuned with respect to natural librational dynamics of the water molecules in their extensive hydrogen-bonded network, which can have an effect on the birefringence as well as on the dichroism.

We note that the transient anisotropy detected in liquid water is characterized by both a fast response at pump-probe overlap ($\ll 5\ ps$) as well as a slower dynamic persisting on the time scale of the macropulse (8 μs). Thus, the signal could possibly be due to entangled nonlinear, thermal, and dissipative effects. We speculate that the fast response originates from the THz torque reorienting the water dipoles via a third-order nonlinear birefringent Kerr effect. The slower dynamic might stem from a combination of several effects such as the formation of a thermal lens, the non-Boltzmann distribution in the unrelaxed HGS, or the alignment of the collective dipole of hydrogen-bonded molecules, because the Debye band of liquid water extends down to 25 MHz, i.e. the repetition rate of the FEL micropulses[108].

We also report the results of a pump-probe signal from a low concentration $MgSO_4$ salt solution which is decreased by ~16-20% with respect to pure water, a value supported both by experiment and theory. This change in direction, i.e. a reduction of the signal, could be understood to arise from the fact that fewer water molecules can participate in the resonantly driven reorientational motions faster than <5 ps imposed by the THz radiation. Because of the stiffer rotational response of the water molecules close to the ions, especially of those waters near the $Mg^{2+}$ cation, their frequency spectrum is blue-shifted and thus could not follow the dynamical reorientation driven by the THz field at 12.3 THz. The pump-probe THz spectroscopy from experiment and theory that we have presented here for water and simple salt solution holds the promise to manipulate and/or map hydration dynamics by the presence of strong AC and DC electric fields. This may open the door to new ways to probe or even control solvation behavior in more complicated environments, e.g. at interfaces and to induce chemical transformations in water by using THz fields.

## ASSOCIATED CONTENT

**Supporting Information:** Pulse overlap and time zero; Nonlinear third-order terms; Cascaded $\chi^{(2)}$ processes; Validation of the AMOEBA force field; Nonlinear optics calculations; Additional control measurements; Estimating the average temperature (diffusive heating).

## AUTHOR INFORMATION

[†]Correspondence to: thg@berkeley.edu

[*]Correspondence to: martina.havenith@rub.de

## Notes

The authors declare no competing financial interests.

## ACKNOWLEDGMENTS

Gefördert durch die Deutsche Forschungsgemeinschaft (DFG) im Rahmen der Exzellenzstrategie des Bundes und der Länder – EXC 2033 – Projektnummer 390677874 – RESOLV. Funded by the Deutsche

Forschungsgemeinschaft (DFG, German Research Foundation) under Germany´s Excellence Strategy – EXC-2033 – Projektnummer 390677874. F. N. and M. H. acknowledge funding by the ERC Advanced Grant 695437. This work is part of the research programme of the Netherlands Organisation for Scientific Research (NWO) and supported by CALIPSOplus (grant agreement no. 730872, EU-H2020). T. H.-G. and L. R. P. thank the U.S. DOE under the Basic Energy Sciences program CPIMS, Contract No. DE-AC02-05CH11231. K. C. B. thanks the California Alliance Postdoctoral Fellowship. T. H.-G. appreciates the support received as a RESOLV Fellow while on sabbatical in Bochum Germany. This research used computational resources of the National Energy Research Scientific Computing Center, a DOE Office of Science User Facility supported by the Office of Science of the U.S. Department of Energy under Contract No. DE-AC02-05CH11231, under an ASCR Leadership Computing Challenge (ALCC) award.

TOC Graphic

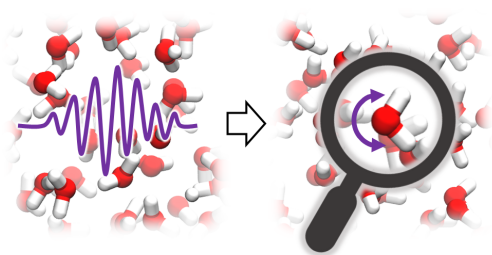